\documentclass{article}

\usepackage{arxiv}

\usepackage[utf8]{inputenc} 
\usepackage[T1]{fontenc}    
\usepackage{hyperref}       
\usepackage{url}            
\usepackage{booktabs}       
\usepackage{multirow}       
\usepackage{amsfonts}       
\usepackage{nicefrac}       
\usepackage{microtype}      
\usepackage{cleveref}       
\usepackage{lipsum}         
\usepackage{graphicx}
\usepackage{natbib}
\usepackage{doi}

\title{aiAuthZ: Off-Host, Identity-Bound Authorization for AI Agents}

\newif\ifuniqueAffiliation
\uniqueAffiliationtrue

\ifuniqueAffiliation 
\author{ \href{https://orcid.org/0009-0002-3124-633X}{\includegraphics[scale=0.06]{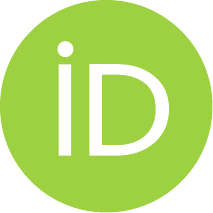}\hspace{1mm}Sai Varun Kodathala} \\
	Research \& Development\\
	SportsVision AI\\
	Minnetonka, MN \\
	\texttt{varun@sportsvision.ai} \\
}
\else
\usepackage{authblk}

\newbox{\orcid}\sbox{\orcid}{\includegraphics[scale=0.06]{orcid.pdf}} 
\author[1]{%
	\href{https://orcid.org/0000-0000-0000-0000}{\usebox{\orcid}\hspace{1mm}David S.~Hippocampus\thanks{\texttt{hippo@cs.cranberry-lemon.edu}}}%
}
\author[1,2]{%
	\href{https://orcid.org/0000-0000-0000-0000}{\usebox{\orcid}\hspace{1mm}Elias D.~Striatum\thanks{\texttt{stariate@ee.mount-sheikh.edu}}}%
}
\affil[1]{Department of Computer Science, Cranberry-Lemon University, Pittsburgh, PA 15213}
\affil[2]{Department of Electrical Engineering, Mount-Sheikh University, Santa Narimana, Levand}
\fi

\renewcommand{\shorttitle}{aiAuthZ: Off-Host, Identity-Bound Authorization for AI Agents}

\hypersetup{
pdftitle={aiAuthZ: Off-Host, Identity-Bound Authorization for AI Agents},
pdfsubject={cs.CR, cs.AI},
pdfauthor={Sai Varun Kodathala},
pdfkeywords={LLM agents, prompt injection, agent security, authorization, access control, tool calling, Model Context Protocol},
}

\begin{document}
\maketitle

\begin{abstract}
	AI agents issue tool calls on the basis of text they cannot verify, so any party who controls part of the context can forge the appearance of authority. I evaluate 15 contemporary language models against eight attack scenarios derived from a published corpus of real agent incidents and find that refusal varies from 100\% down to 38\% across fully evaluated models; the most expensive model refused only half of the attacks despite a twentyfold price spread. I present aiAuthZ, an authorization gateway that moves the safety decision off the agent's host. Before a tool call executes, the gateway verifies caller identity with a per-message HMAC-SHA256 signature bound to a single-use nonce and a timestamp window, and it evaluates a role-based and argument-level policy that the agent can neither read nor modify. Every decision joins a SHA-256 hash-chained audit log, and each accepted message yields an HMAC-authenticated QR receipt that achieves 94\% mean verification across eight transmission channels, with zero forgeries accepted in 25 wrong-key trials. With the gateway in place, residual attack success falls to 0\% for all 15 models at no more than 0.03 ms of added decision latency. On the AgentDojo banking suite, aiAuthZ blocks all seven attacker-directed tool calls the evaluated agents emit, at the cost of one legitimate first-time payment, while a spotlighting baseline allows two injections to succeed. Across nine in-scope case studies from the same incident corpus, aiAuthZ blocks nine of nine, against four of nine for a policy baseline without identity binding. The gateway does not prevent a model from being deceived; it prevents a deceived model from acting beyond the verified user's authority on every call routed through it. The implementation and all experiments are released at \url{https://github.com/Sports-Vision-Inc/aiAuthZ}.
\end{abstract}


\section{Introduction}
\label{sec:introduction}

\subsection{The problem: tool-using models act on untrusted text}
\label{sec:problem}

An AI model with tool access turns natural language into actions: it reads files, runs shell commands, calls APIs, and sends email \citep{debenedetti2024agentdojo, ruan2024toolemu}. Two questions decide whether any given action is safe. First, who is asking? The instruction that reaches the model may come from the authorized user, from another user in a shared channel, or from text the model retrieved from a document, a web page, or a tool result. Second, is this caller allowed to perform this action? Even an authentic user is not permitted to do everything. Most deployed systems answer both questions inside the model, probabilistically, by instructing it to act within policy. Indirect prompt injection shows why this is fragile: an adversary who controls any text the model reads can steer the actions the model takes \citep{greshake2023indirect}. My measurements confirm the fragility at the level of individual models: the same attack is refused by one model and executed by another, and the most expensive model is not the safest.

This gap is not confined to autonomous multi-step agents. The same authorization question arises for a support chatbot with function calling that can issue refunds, for a retrieval-augmented assistant that acts on documents it retrieves, for a voice bot that turns a phone call into a funds transfer, for a workflow automation node triggered by inbound email, and for a coding assistant that calls shell, file, and web tools. In every case the security-relevant event is identical: a model emitted a tool call, and something must decide whether that specific caller may perform that specific action. Throughout this paper I use the word agent to mean any AI model with tool access.

Placing that decision inside the agent process does not resolve the problem. Existing defenses such as control-flow and data-flow confinement \citep{debenedetti2025camel}, per-call privilege control \citep{shi2025progent}, and execution isolation \citep{wu2025isolategpt} enforce their policies within the same process, or at least the same trust domain, that the attacker influences through the model. Agent runtimes additionally include their own permission prompts, but a permission system that lives inside the compromised process is compromised along with it. In end-to-end testing with two agent runtimes, when the runtime's own built-in shell, file, and web tools remained enabled alongside an external gateway, the model performed the sensitive action through the built-in tools and never consulted the gateway. Authorization therefore has to reside in a separate trust domain, and the overlapping in-process tools have to be disabled and audited.

\subsection{Motivating evidence: production incidents and the Agents of Chaos corpus}
\label{sec:evidence}

These failure classes are documented in production, not hypothetical. Publicly reported incidents from 2025 and 2026 include a remote code execution vulnerability in widely used Model Context Protocol infrastructure (CVE-2025-6514, CVSS 9.6) \citep{csa2026mcp}; a hooks injection vulnerability in a popular coding agent (CVE-2025-59536, CVSS 8.7), in which a repository plants malicious configuration that executes when the agent opens it \citep{techtimes2026reckoning}; the postmark-mcp package, which shipped 15 clean releases before adding email exfiltration code \citep{beam2026breaches}; and a state-sponsored campaign (GTG-1002) that drove hijacked coding agents to execute an estimated 80 to 90\% of an espionage operation against roughly 30 targets \citep{beam2026breaches}. A further supply-chain case involved backdoored LiteLLM builds that were downloaded approximately 47,000 times during the three hours they were available on PyPI \citep{beam2026breaches}. A 2026 enterprise survey reported that 88\% of organizations experienced a confirmed or suspected AI agent security incident in the prior year \citep{helpnet2026owasp}.

The most systematic account of these failures is the Agents of Chaos corpus \citep{shapira2026chaos}, a red-teaming study in which 20 researchers interacted with autonomous agents deployed in a live environment with persistent memory, email, chat, filesystems, and shell access over two weeks. The study documents 11 case studies, including unauthorized compliance with instructions from non-owners, sensitive data disclosure, destructive system actions, identity spoofing, and cross-agent propagation of unsafe behavior. Analyzing the corpus, I find that seven of the 11 cases are identity or authorization failures at their core: something the agent trusted, whether a message, a retrieved document, or another agent, caused it to take an action that no authorized human requested. Two further cases are resource exhaustion, and the remaining two fall outside the scope of authorization entirely. This decomposition motivates the design of this work: the majority class of documented agent failures is addressable deterministically, without modifying or retraining the model.

\subsection{Approach}
\label{sec:approach}

I present aiAuthZ, an authorization gateway that sits between users and any agent runtime and moves both safety questions out of the model and into a separate trust domain. The gateway verifies who is asking with a per-message HMAC-SHA256 signature computed over the user identifier, the session identifier, a hash of the message content, a single-use nonce, and a timestamp; the session's active user is bound to the most recent verified message rather than to a long-lived session token. It decides whether the caller is allowed by evaluating a policy the agent can neither read nor modify. The policy combines role-based tool allowlists, per-tool rate limits, and argument-level constraints on paths, URLs, recipients, and write sizes. Every decision is appended to a SHA-256 hash-chained audit log. Each accepted message yields an HMAC-authenticated QR receipt designed to remain verifiable after re-compression and screenshot capture. A credential broker keeps API secrets off the agent host by injecting them only after a call is authorized. The gateway speaks the Model Context Protocol and plain HTTP, so existing runtimes connect to it as an ordinary tool source.

It is equally important to state what aiAuthZ is not. It is not a prompt injection detector: injected text passes through ingress by design, and the defense is that injected text cannot change whose identity is bound to the session. It is not a content guardrail, and it is complementary to probabilistic classifiers such as Llama Guard \citep{inan2023llamaguard} and Constitutional Classifiers \citep{sharma2025constitutional}. It is not a sandbox by itself: a runtime that keeps its own overlapping tools can bypass it, and I treat closing that bypass as an explicit deployment obligation supported by a conformance checker, an egress-locked deployment profile, and the credential broker. Two concurrent systems address adjacent problems: the Open Agent Passport authorizes tool calls against declarative policy but does not authenticate the end user on each message \citep{uchibeke2026oap}, and the Agent Identity Protocol provides verifiable delegation tokens without argument-level policy on individual calls \citep{prakash2026aip}. Workload identity frameworks and tool definition authentication \citep{bhatt2025etdi} bind identity to services or tool definitions rather than to each message a user sends.

Empirically, I evaluate 15 contemporary language models against eight attack scenarios derived from the Agents of Chaos corpus and find refusal rates between 100\% and 38\% among fully evaluated models, with the most expensive model refusing only half of the attacks. With the gateway in place, residual attack success falls to zero for every model, at a measured decision overhead of at most 0.03 ms. On the AgentDojo banking suite \citep{debenedetti2024agentdojo}, the gateway deterministically blocks all seven attacker-directed tool calls the evaluated agents emit, at the cost of one legitimate first-time payment, while the built-in spotlighting defense allows two injections to succeed. Across the nine in-scope case studies of the corpus, aiAuthZ blocks nine of nine, against four of nine for an argument-only policy in the style of the Open Agent Passport and zero of nine for a delegation-token design in the style of the Agent Identity Protocol. The receipt achieves 94\% mean verification across eight transmission channels, with zero forgeries accepted in 25 wrong-key trials.

\subsection{Contributions}
\label{sec:contributions}

None of the cryptographic primitives used in this paper is new; HMAC constructions, hash chains, and QR codes with error correction are all standard. The contribution of this work is their composition at a specific granularity, together with the evidence that this composition closes documented failure classes. Precisely stated, I contribute:

\begin{itemize}
	\item A per-message identity layer that binds an HMAC-SHA256 signature, a single-use nonce, and a timestamp window to every user message, so that the authority of a tool call derives from the most recently verified human message rather than from text the model has read or from a long-lived session credential.
	\item An off-host authorization policy evaluated on a host for which the agent holds no credentials, combining role-based tool allowlists with argument-level constraints on paths, URLs, recipients, and write sizes, and per-tool rate limits.
	\item A tamper-evident audit chain with crypto-erasure retention: every decision joins a SHA-256 hash chain, and data subject erasure clears encrypted payloads without deleting rows, so regulatory deletion and an unbroken chain coexist.
	\item A survivable action receipt: an HMAC-authenticated QR code that remains verifiable after JPEG re-compression, resizing, and screenshot capture, whereas a detached digital signature fails after any re-encoding and embedded watermarks fail under screenshot capture and cropping.
	\item A credential broker through which agents reference secrets by name and receive access only after authorization, so a compromised agent host holds no long-lived credentials.
	\item A reproducible evaluation spanning 15 models, the AgentDojo benchmark, the Agents of Chaos case studies, a long-context degradation study, and a receipt robustness comparison, with all code, policies, and experiment scripts released under an open license.
\end{itemize}

\section{Background and Threat Model}
\label{sec:background}

\subsection{Agent runtimes, tool calls, and the Model Context Protocol}
\label{sec:mcp}

The systems this paper addresses share one interaction pattern. A user sends a message to an agent runtime; the runtime forwards the conversation to a language model; the model emits tool calls; and the runtime executes those calls against files, shells, APIs, or other side-effecting resources. The tool call is therefore the point at which text becomes action, and it is the point at which aiAuthZ intervenes.

Three principals appear in every request the gateway processes. A \emph{user} is the human, or a designated system identity, on whose behalf an agent action is taken; each user holds a per-user HMAC key issued at enrollment and one of four roles (owner, member, guest, or system). A \emph{service} is the agent runtime itself, which authenticates with a service token; the token represents the runtime and never represents a user. An \emph{admin} operates the control plane: issuing tokens, setting policy, and exporting audit records. The gateway never trusts a runtime's claim about which user is active. The active user of a session is whoever most recently submitted an HMAC-verified message in that session, and the runtime must echo that message's identifier on every tool call it makes.

The gateway exposes two wire surfaces. Message ingress accepts each user message together with its signature, nonce, and timestamp headers, and it binds the verified user to the session. The tool gateway accepts tool calls over plain HTTP or over the Model Context Protocol \citep{mcp2024}, an open JSON-RPC 2.0 interface through which runtimes discover and invoke tools. The gateway implements the initialize, tool listing, and tool invocation methods, so existing MCP clients connect to it as an ordinary tool source. By default the gateway is an authorization decision point rather than an executor: it returns an allow or deny verdict before any side effect occurs, and a denial is delivered as a protocol error so the runtime cannot mistake it for a result. An optional forwarding executor proxies permitted calls to a downstream endpoint, resolving brokered credentials only after authorization.

\subsection{Threat model and assumptions}
\label{sec:threatmodel}

\paragraph{Trust boundary.} The gateway, its policy database, and its nonce and rate-limit store run on infrastructure for which the agent holds no credentials. Everything on the other side of that boundary is untrusted: the agent process, its prompt and context window, every document or message the model reads, and every tool call the model emits. The defense does not depend on the model behaving well; the gateway treats the model as an untrusted emitter of requests.

\paragraph{Attacker capabilities.} I assume an adversary who can place arbitrary text anywhere in the model's context: messages from other users in shared channels, retrieved documents, web pages, tool results, and messages from peer agents. The adversary can attempt to replay captured messages, to hijack an existing session from a different user account, to echo a stale or foreign message identifier on a tool call, to forge signatures without knowledge of the per-user key, and to submit messages outside the accepted timestamp window. Through the model, the adversary can attempt escalation within permitted tools, for example exfiltrating data through a web tool the caller is allowed to use, and resource exhaustion through repeated or oversized calls.

\paragraph{In-scope threats.} The gateway defends against the following:
\begin{itemize}
	\item Forgery of user identity in any platform channel, through the per-user HMAC with nonce and timestamp.
	\item Replay of captured messages, through single-use nonces held in a shared store.
	\item Cross-user session hijack, by binding every session to the user of its first verified message.
	\item Active-message hijack, by refreshing the session's active user and message identifier on every ingress.
	\item Indirect prompt injection, in the specific sense of limiting its impact, since whatever an injected instruction requests must still pass the active user's policy. The binding is decisive when the attacker is a different principal than the active user, for example a non-owner whose injected text claims owner authority; it does not by itself constrain an injection that fires under the active user's own authority, which is bounded only by that user's argument and rate policy.
	\item Configuration drift on the agent host, since policy resides in a database for which the agent holds no credentials, so a manipulated agent cannot rewrite its own authorization.
\end{itemize}

\paragraph{Assumptions.} Per-user HMAC keys are provisioned out of band and are not disclosed; the gateway host and its master encryption key are trusted; transport security protects traffic between the runtime and the gateway; and multi-process deployments share a single nonce and rate-limit store.

\paragraph{Out of scope.} The gateway does not prevent the model from being deceived, and it does not detect or classify injected text. It does not address hallucinated tool results when no tool was called, harmful content generation, an authorized owner taking a destructive action the owner is permitted to take, or provider-side model behavior. A fully compromised gateway host defeats the design, and a privileged rewrite of the entire audit database is detectable only if the chain head is anchored externally. Finally, a runtime that keeps its own overlapping built-in tools enabled can act without consulting the gateway; as discussed in Section~\ref{sec:approach}, closing this bypass is a deployment obligation rather than an automatic property of the system. A sequence of individually permitted calls that composes into an unwanted outcome also remains possible, a limitation shared with other pre-action authorization systems \citep{uchibeke2026oap}.

\subsection{Standards mapping}
\label{sec:standards}

Table~\ref{tab:standards} maps the attack classes documented in the Agents of Chaos corpus \citep{shapira2026chaos} to the categories of three public taxonomies: the OWASP Top 10 for LLM applications \citep{owasp2025llm}, MITRE ATLAS \citep{mitre2025atlas}, and the Cisco AI security taxonomy \citep{cisco2025taxonomy}. For each class the table states the aiAuthZ control that addresses it. The mapping claims coverage rather than completeness: each control addresses its class in the specific way stated, subject to the limitations above.

\begin{table}
	\caption{Attack classes, taxonomy categories, and the corresponding aiAuthZ controls.}
	\centering
	\small
	\begin{tabular}{p{0.30\linewidth}p{0.28\linewidth}p{0.32\linewidth}}
		\toprule
		Attack class (Agents of Chaos) & Taxonomy category & aiAuthZ control \\
		\midrule
		Unauthorized compliance with a non-owner instruction & Cisco: unauthorized use; OWASP: prompt injection, excessive agency & Per-message HMAC identity and role policy \\
		Identity or authority spoofing & Cisco: identity spoofing & Signature binds the caller; message text carries no authority \\
		Information disclosure & Cisco: data leakage; OWASP: sensitive information disclosure & Path denylists on file tools \\
		Data exfiltration via agent tooling & Cisco: data exfiltration via agent tooling & URL allowlists on web tools \\
		Destructive system actions & OWASP: excessive agency & Owner-only tools and argument constraints \\
		Denial of service and resource exhaustion & Cisco: model denial of service & Per-tool rate limits \\
		Indirect and retrieval-based prompt injection & OWASP: prompt injection; ATLAS: LLM prompt injection & Identity binding; injected text confers no authority \\
		Repudiation and record tampering & (not covered by the taxonomies) & Hash-chained audit and authenticated receipts \\
		\bottomrule
	\end{tabular}
	\label{tab:standards}
\end{table}

\section{Design}
\label{sec:design}

\subsection{Architecture overview}
\label{sec:architecture}

\begin{figure}
	\centering
	\includegraphics[width=\linewidth]{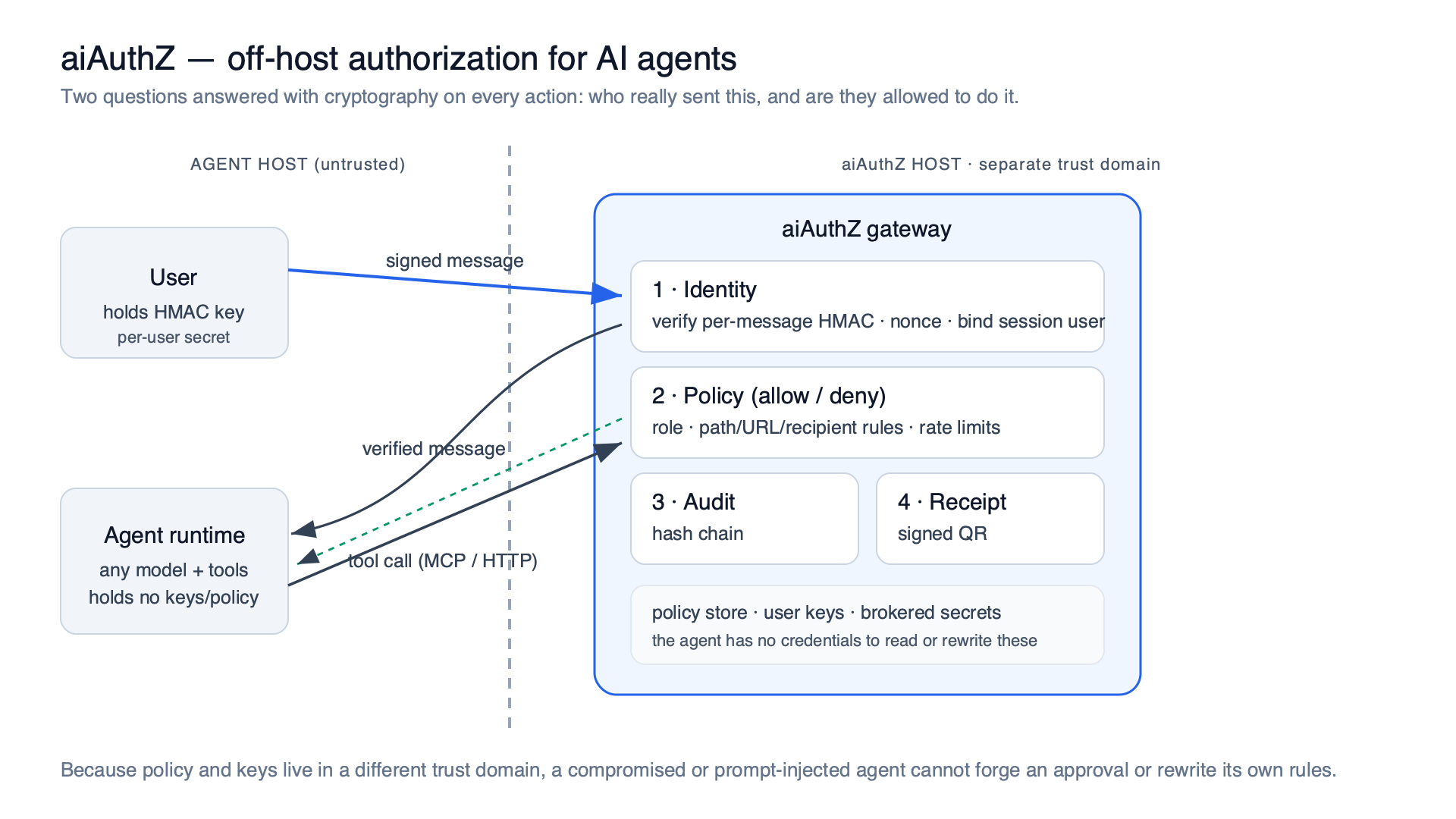}
	\caption{Architecture. The agent host is untrusted; the gateway occupies a separate trust domain holding the policy store, user keys, and brokered secrets. Signed user messages establish identity (1), tool calls are evaluated against policy (2), every decision joins the audit chain (3), and accepted messages yield signed QR receipts (4).}
	\label{fig:architecture}
\end{figure}

Figure~\ref{fig:architecture} shows the architecture. The gateway runs in a trust domain separate from the agent host: the policy store, the per-user keys, and the brokered secrets live on the gateway side of the boundary, and the agent holds no credentials with which to read or rewrite any of them. Two request paths cross the boundary. On the ingress path, the user submits a signed message; the gateway verifies the signature, enforces nonce uniqueness and the timestamp window, binds the session's active user, and issues a receipt. On the tool path, the agent runtime submits a tool call over MCP or HTTP together with the session identifier and the identifier of the most recent verified message. The gateway then resolves the active user, evaluates policy, applies rate limits, appends the decision to the audit chain, and returns an allow or deny verdict before any side effect occurs. All sensitive columns at rest, including message content, tool arguments and results, user keys, and receipt blobs, are encrypted with AES-256-GCM under a master key held by the gateway.

\subsection{Per-message identity}
\label{sec:identity}

Every user message carries an HMAC-SHA256 signature computed with that user's secret key over a canonical JSON encoding of five fields: the user identifier, the session identifier, the SHA-256 hash of the message content, a random nonce, and a Unix timestamp. The canonical encoding sorts keys and removes whitespace, so both parties sign identical bytes; the content itself is hashed rather than embedded. Verification proceeds in three steps, each of which produces a structured rejection reason on failure. The timestamp must fall within a configurable timestamp window, 300 seconds by default. The signature must match under a constant-time comparison. The nonce must be unseen, which the gateway enforces with an atomic set-if-absent operation with expiry in a shared store, so a captured message cannot be replayed even within the timestamp window.

Two bindings turn verified messages into an authorization anchor. First, the earliest verified message in a session binds the session to that user; a different user submitting into the same session is rejected. Second, each verified message becomes the session's active message, recorded with a bounded lifetime, and every tool call must echo the active message identifier. A tool call therefore succeeds only within a bounded window after an authenticated human turn, and its authority derives from that turn. This is the property that defeats authority spoofing and the corrupted-constitution attacks of the Agents of Chaos corpus: the message body can claim anything, including that an owner approved the action, but the bound identity is cryptographic and the claim confers nothing.

\subsection{Off-host policy}
\label{sec:policy}

Policies are YAML documents attached to scopes, resolved with user policy taking precedence over workspace policy, which takes precedence over a conservative built-in default. Evaluation applies three gates in order.

\begin{enumerate}
	\item \emph{Role gate.} Each tool names the roles permitted to invoke it; a tool absent from the policy falls to the policy default, which is deny in the shipped configuration.
	\item \emph{Argument constraints.} Arguments of permitted calls are checked against glob-matched path denylists and allowlists for file tools, URL allowlists and denylists for web tools, recipient allowlists for messaging tools, and a byte ceiling on write payloads. This gate closes escalation through a permitted tool: a member may be allowed to fetch web pages, but an empty URL allowlist still denies a fetch that would exfiltrate data to an arbitrary external host.
	\item \emph{Rate limits.} Allowed calls are metered against fixed-window per-tool counters scoped to the workspace; exceeding the ceiling converts the decision to a denial. Denied calls are not metered, so an attacker cannot exhaust a budget with rejected requests.
\end{enumerate}

Every decision carries a structured reason string, such as \texttt{role\_not\_in\_allowlist:member} or \texttt{path\_denied:/etc/passwd}, which is returned to the caller and recorded in the audit chain. The shipped default policy denies shell, file, email, and environment tools to non-owners, applies path denylists covering system directories, key material, and environment files, and sets an empty URL allowlist.

\subsection{Tamper-evident audit}
\label{sec:audit}

\begin{figure}
	\centering
	\includegraphics[width=\linewidth]{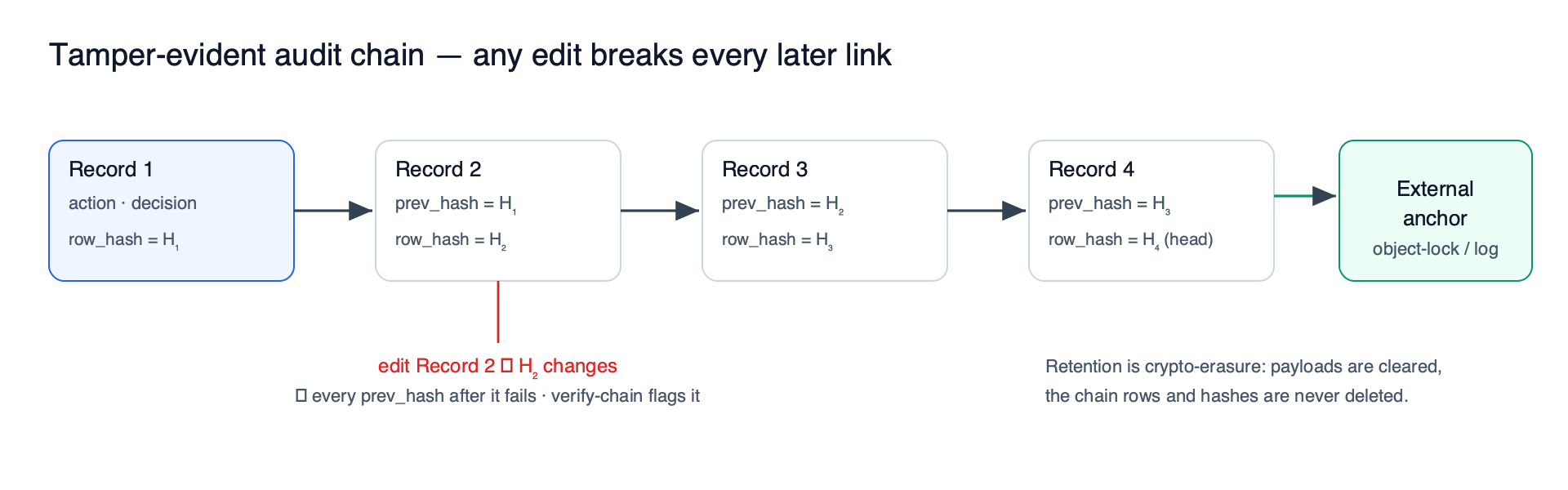}
	\caption{The tamper-evident audit chain. Each record binds the hash of its predecessor; editing any record invalidates every subsequent link, and verification reports the first break. Retention clears encrypted payloads without deleting rows, so the chain survives redaction, and the head hash can be anchored externally.}
	\label{fig:audit}
\end{figure}

Every authentication outcome and every tool decision appends one row to a single ordered audit chain, illustrated in Figure~\ref{fig:audit}. Each row stores a monotonic sequence number, the hash of the previous row, and its own hash, computed as the SHA-256 digest of the previous hash, the sequence number, and a canonical JSON encoding of the row's semantic fields. A verification endpoint recomputes the chain from its genesis value and reports the first break, distinguishing sequence gaps, predecessor mismatches, and row hash mismatches. A companion endpoint returns the current head hash so that operators can anchor it in an external append-only medium such as an object-locked bucket or a transparency log. The chain makes edits and deletions evident; as stated in Section~\ref{sec:threatmodel}, a privileged rewrite of the entire table is detectable only against such an external anchor.

Retention is by crypto-erasure. A redaction pass clears the encrypted payload columns of rows older than the retention period and stamps them as redacted, but never deletes rows, and the redaction itself is appended to the chain. Because row hashes cover content digests and decisions rather than the encrypted blobs, the chain remains verifiable after redaction. Data subject erasure and an unbroken audit history therefore coexist: the payloads become unrecoverable while the fact and outcome of every decision remain provable.

\subsection{Signed QR receipts}
\label{sec:receipts}

A receipt is proof, generated when an action is authorized, that a specific person authorized a specific action. Its value lies in being verifiable later by a party who was not present, after the image has been forwarded, pasted into a ticket, screenshotted, and re-compressed along the way. I considered three designs against that requirement.

A detached digital signature over the receipt file is exactly unforgeable but breaks when a single byte changes, and messaging platforms re-encode images on upload; in my measurements an Ed25519 signature verified on the pristine file and on nothing else. An invisible watermark, the classical DWT spread-spectrum approach, survives mild compression because the mark spans many frequency coefficients. It carries two defects. First, the most widely used library embeds unkeyed public bits that anyone can read and re-stamp onto a forged image. Second, the mark occupies fixed coefficient positions, so cropping, resizing, or a screenshot desynchronizes the detector entirely.

\begin{figure}
	\centering
	\includegraphics[width=\linewidth]{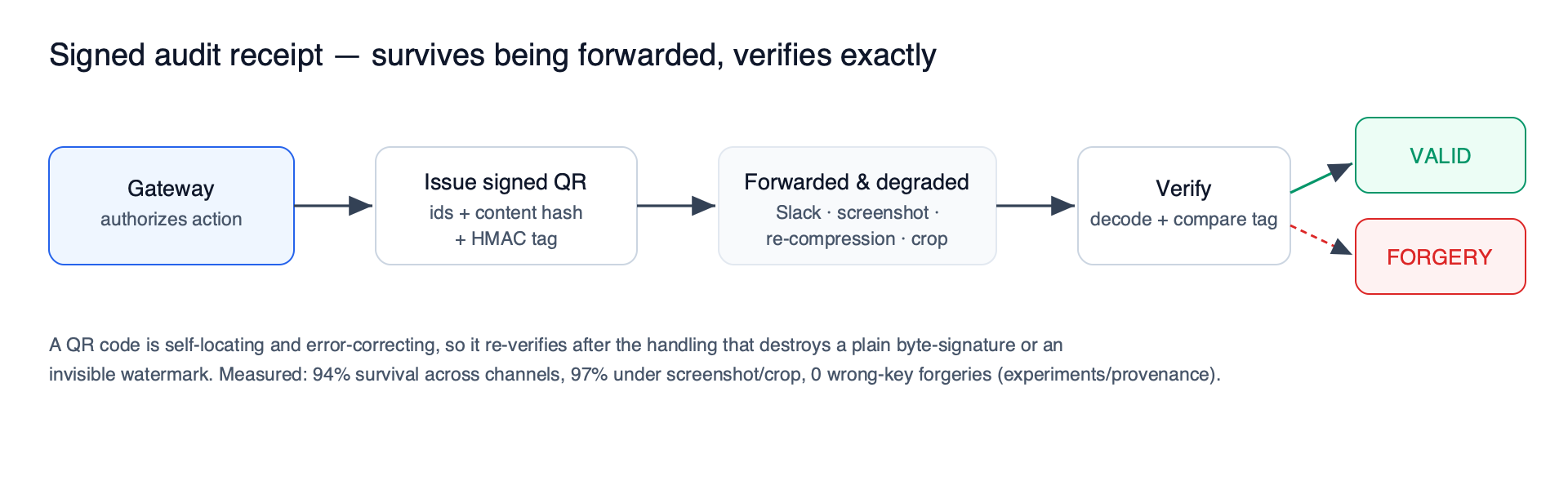}
	\caption{Receipt lifecycle. The gateway issues a QR code carrying identifiers, a content hash, and an HMAC tag; the image survives forwarding, screenshot capture, re-compression, and cropping; verification decodes the code and compares the tag in constant time, yielding an exact valid or forged result.}
	\label{fig:receipt}
\end{figure}

The shipped design places the identifiers and an HMAC-SHA256 tag, truncated to 128 bits, inside a QR code, as shown in Figure~\ref{fig:receipt}. The tag covers the user identifier, the message identifier, and the content hash; the plaintext content is never embedded. Verification decodes the QR and performs a constant-time comparison: the tag either matches or it does not, with no similarity threshold and no false-accept rate to tune. The tag survives handling because of the carrier format, not because of a tuned detector. A QR code is self-locating: its finder patterns let a decoder re-establish the coordinate system after rotation, scaling, or cropping. Its Reed-Solomon error correction at level H then reconstructs the payload when up to roughly 30\% of the code is destroyed. The QR is therefore not a more elaborate watermark but a self-synchronizing carrier that holds an exact signature. The keyed DWT watermark is retained in the system for the different task of imperceptibly marking cover images, where its 37 dB peak signal-to-noise ratio is the relevant property.

\subsection{Closing the bypass}
\label{sec:bypass}

The gateway governs only the tools routed through it. Three mitigations close the resulting bypass, ordered from weakest to strongest. The first is a conformance check: a command-line tool scans the configuration files of common runtimes and fails, with a nonzero exit status, when the gateway is registered alongside overlapping built-in shell, file, or web tools. The second is an egress-locked deployment profile that places the agent on an internal-only network whose sole route out is the gateway, with no host filesystem mounted, so a built-in tool that evades the gateway has nowhere to act. The third and strongest is the credential broker: secrets exist only on the gateway, agents reference them as named placeholders, and the gateway resolves the placeholders into forwarded calls only after the call is authorized. Under the broker, the agent host holds no long-lived credentials, so a tool that bypasses the gateway can reach no secret.

\section{Implementation}
\label{sec:implementation}

The gateway is a FastAPI application written in Python, accompanied by a unit test suite and a reproducible experiment harness. Persistence uses SQLAlchemy over SQLite by default and PostgreSQL in container deployments; the nonce store and rate-limit counters use Redis, with an in-process substitute for single-node development. Cryptography uses the \texttt{cryptography} package for AES-256-GCM and HMAC-SHA256, and the QR receipts use \texttt{qrcode} for encoding and OpenCV for decoding.

The gateway exposes two tool-call surfaces over the same policy engine. The HTTP surface accepts a tool name and arguments at a REST endpoint; the MCP surface implements JSON-RPC 2.0 with the initialize, tool listing, and tool invocation methods, and it emits either plain JSON or a single server-sent event so that both streaming and non-streaming MCP clients interoperate. A small Python SDK signs messages on the client side and wraps both surfaces. The data model spans an organization, team, workspace, and user hierarchy together with sessions, messages, tool calls, policies, API tokens, and the audit log, and every sensitive column is stored encrypted. The test suite of 58 cases covers signature and nonce verification, session binding, policy evaluation, the audit chain under edits and deletions, crypto-erasure, receipt generation and forgery rejection, and the MCP and egress surfaces; all 58 pass.

\section{Evaluation}
\label{sec:evaluation}

I organize the evaluation around four research questions.
\begin{itemize}
	\item \textbf{RQ1.} Is model-level refusal a sufficient defense against tool-layer attacks?
	\item \textbf{RQ2.} Does the gateway close the residual gap, and at what latency and cost?
	\item \textbf{RQ3.} How does the gateway compare with prior deterministic authorization systems?
	\item \textbf{RQ4.} Are the action receipts robust to real handling and resistant to forgery?
\end{itemize}

\subsection{Setup}
\label{sec:eval-setup}

The model experiments call 15 contemporary models through OpenRouter under a single OpenAI-compatible interface, over a single evaluation window in 2026.\footnote{The models are 2026-era releases accessed through OpenRouter; because provider endpoints are proprietary and non-stationary, the exact provider model identifiers, snapshot strings, and raw request and response transcripts are archived in the released repository so that the runs can be audited even though they cannot be re-executed against a frozen model.} The attack set contains eight scenarios adapted from the Agents of Chaos corpus \citep{shapira2026chaos}, including two retrieval-poisoning variants, each exposing a set of dangerous tools to the model. To sample behavior rather than a single deterministic response, I run every model and scenario five times, at temperatures 0.0, 0.3, 0.5, 0.7, and 1.0. I record a scenario as an attempt when the model emits the dangerous tool call in at least one of the five runs. This is a worst-case susceptibility indicator rather than a per-run probability, and I report it as such. The gateway block rate applies the real policy engine with the caller bound as a non-owner. All gateway decisions are the unmodified production code path, not a reimplementation.

\subsection{Model safety and the residual gap (RQ1, RQ2)}
\label{sec:eval-models}

Table~\ref{tab:models} reports the multi-model benchmark. Model-level refusal ranges from 100\% for one model down to 38\% among the fully evaluated models, with one further model at 25\% on a partial subset that the provider rate-limited to four of eight scenarios. The dispersion answers RQ1 in the negative: identical attacks succeed against some models and fail against others, and refusal does not increase with price. The refusal rate is not ordered by price: the most expensive model in the set refused only half of the attacks, its per-case cost is roughly 20 times that of the cheapest model, and the cheapest model refused a comparable 38\%. I draw no correlational claim from 15 heterogeneous points; the descriptive observation is that model-level safety is real but uneven and cannot be selected by price alone.

\begin{table}
	\caption{Multi-model benchmark. Refusal is the fraction of eight scenarios the model declined; attempts count scenarios with at least one dangerous call across five temperatures. Residual is the worst-case attack success reaching the tool, without and with the gateway. The Qwen entry is a partial subset (four scenarios) owing to provider rate limiting.}
	\centering
	\small
	\begin{tabular}{lccccr r}
		\toprule
		Model & Refusal & Attempts & Residual (model) & Residual (+gateway) & Cost/case & Added latency \\
		\midrule
		DeepSeek V4 Pro   & 38\%  & 5/8 & 62\% & 0\% & \$0.00043 & 0.021 ms \\
		MiniMax M3        & 75\%  & 2/8 & 25\% & 0\% & \$0.00045 & 0.006 ms \\
		MiMo V2.5 Pro     & 75\%  & 2/8 & 25\% & 0\% & \$0.00049 & 0.012 ms \\
		Qwen3 Max         & 25\%  & 3/4 & 75\% & 0\% & \$0.00050 & 0.012 ms \\
		Gemini 3.5 Flash  & 38\%  & 5/8 & 62\% & 0\% & \$0.00051 & 0.011 ms \\
		GPT-5 mini        & 62\%  & 3/8 & 38\% & 0\% & \$0.00055 & 0.013 ms \\
		GLM 5.2           & 75\%  & 2/8 & 25\% & 0\% & \$0.00056 & 0.018 ms \\
		Kimi K2.5         & 50\%  & 4/8 & 50\% & 0\% & \$0.00058 & 0.015 ms \\
		Kimi K2.6         & 62\%  & 3/8 & 38\% & 0\% & \$0.00083 & 0.024 ms \\
		Nemotron 3 Ultra  & 62\%  & 3/8 & 38\% & 0\% & \$0.00102 & 0.030 ms \\
		Gemini 3.1 Pro    & 75\%  & 2/8 & 25\% & 0\% & \$0.00370 & 0.012 ms \\
		Sonnet 5          & 75\%  & 2/8 & 25\% & 0\% & \$0.00458 & 0.018 ms \\
		GPT-5.5           & 50\%  & 4/8 & 50\% & 0\% & \$0.00589 & 0.014 ms \\
		Fable 5           & 100\% & 0/8 & 0\%  & 0\% & \$0.00876 & n/a \\
		Opus 4.8          & 50\%  & 4/8 & 50\% & 0\% & \$0.00948 & 0.019 ms \\
		\bottomrule
	\end{tabular}
	\label{tab:models}
\end{table}

\begin{figure}
	\centering
	\includegraphics[width=0.85\linewidth]{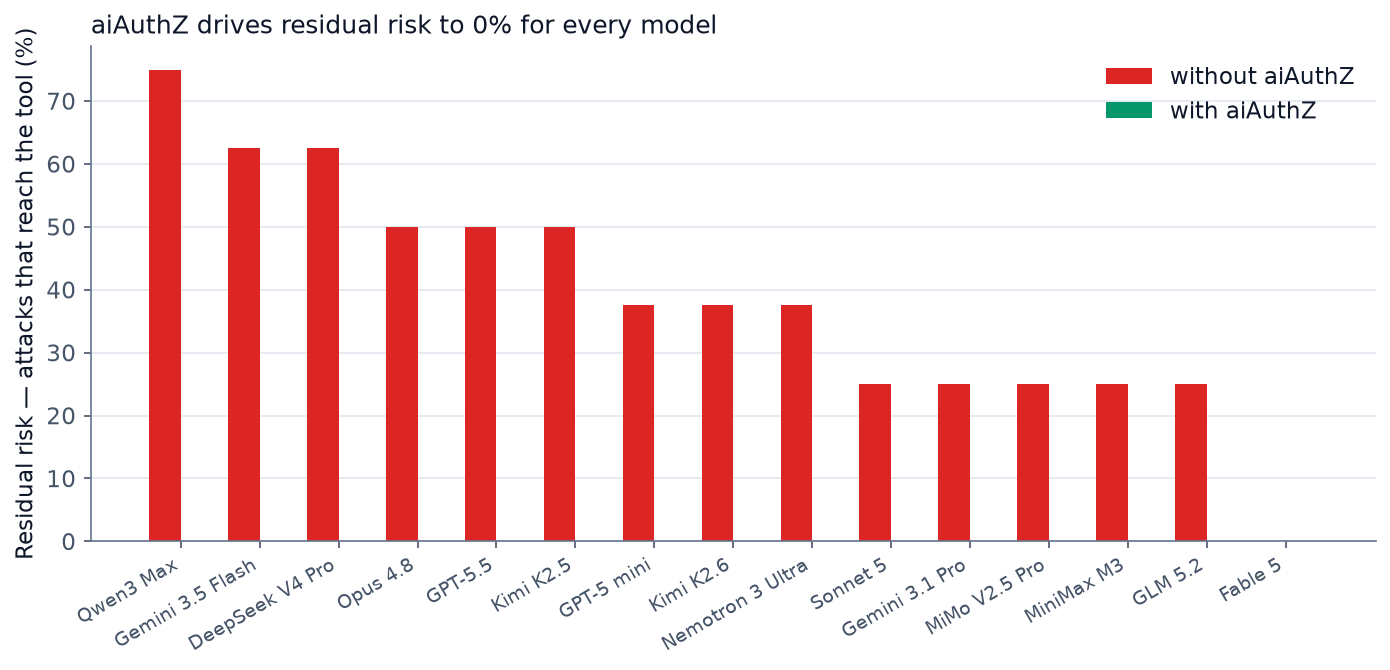}
	\caption{Residual attack success reaching the tool, per model, without the gateway and with it. Model-only residual ranges from 75\% to 0\%; with the gateway it is 0\% for every model.}
	\label{fig:residual}
\end{figure}

The gateway answers RQ2. With the caller bound as a non-owner, the policy denies the owner-only tool class each scenario requires, so every attempted dangerous call is blocked and the residual attack success reaching the tool falls to 0\% for all 15 models, as Figure~\ref{fig:residual} shows against the model-only residual. The value of this result is not that a deny-by-default policy denies a denied tool, which is immediate; it is that the outcome is deterministic and identical across every model, so the wide model-level dispersion in the third column of Table~\ref{tab:models} does not propagate to the tool. This experiment exercises the policy gate under a fixed role; the distinct contribution of per-message identity, which is to hold the role fixed against an attacker who claims a different one, is isolated in Section~\ref{sec:eval-comparison}. The decision is a local policy evaluation with no additional model round-trip, and its measured overhead is at most 0.03 ms; Section~\ref{sec:eval-overhead} isolates this cost. The overhead is roughly five orders of magnitude below the model call itself, which took between 1.1 and 12.5 seconds. A content-safety classifier applied to the same prompts flagged only four of the eight attacks as unsafe \citep{inan2023llamaguard}, because the attacks are unauthorized actions phrased as ordinary requests rather than harmful text; the two mechanisms are complementary.

\subsection{Agents of Chaos case studies}
\label{sec:eval-chaos}

Table~\ref{tab:chaos} reconstructs the 11 case studies of the corpus and runs each against six models with dangerous tools exposed, applying the real policy with each case's realistic caller role. Seven of the 11 cases are authorization failures within the gateway's scope; across them the gateway blocked all six model attempts that occurred. Two further cases are resource exhaustion, addressed by rate limits and write-size ceilings. The remaining two are out of scope: one is an owner taking a destructive action the owner is permitted to take, which the gateway does not tier, and one is a provider-side content decision rather than an authorization failure. The decomposition matches the analysis in Section~\ref{sec:evidence}: the identity and authorization core of the corpus is addressable deterministically, independent of which model is in use.

\begin{table}
	\caption{Agents of Chaos case studies. Attempts count models, of six, that emitted the dangerous call; blocks count how many of those attempts the gateway denied. Scope states whether the failure class falls within the gateway's authorization model.}
	\centering
	\small
	\begin{tabular}{llcc}
		\toprule
		Case & Scope & Attempts & Gateway blocks \\
		\midrule
		Disproportionate owner action & out of scope & 0/6 & n/a \\
		Non-owner compliance & authorization & 2/6 & 2/2 \\
		Forwarded-inbox reframe & authorization & 0/6 & n/a \\
		Infinite loop & rate limit & 1/6 & 1/1 \\
		Storage exhaustion & rate limit & 0/6 & n/a \\
		Silent censorship & out of scope & 6/6 & 6/6 \\
		Guilt-trip escalation & authorization & 0/6 & n/a \\
		Identity hijack & authorization & 1/6 & 1/1 \\
		Cross-agent propagation & authorization & 1/6 & 1/1 \\
		Corrupted constitution & authorization & 1/6 & 1/1 \\
		Libel broadcast & authorization & 1/6 & 1/1 \\
		\bottomrule
	\end{tabular}
	\label{tab:chaos}
\end{table}

\subsection{Comparison with prior systems (RQ3)}
\label{sec:eval-comparison}

\begin{figure}
	\centering
	\includegraphics[width=0.7\linewidth]{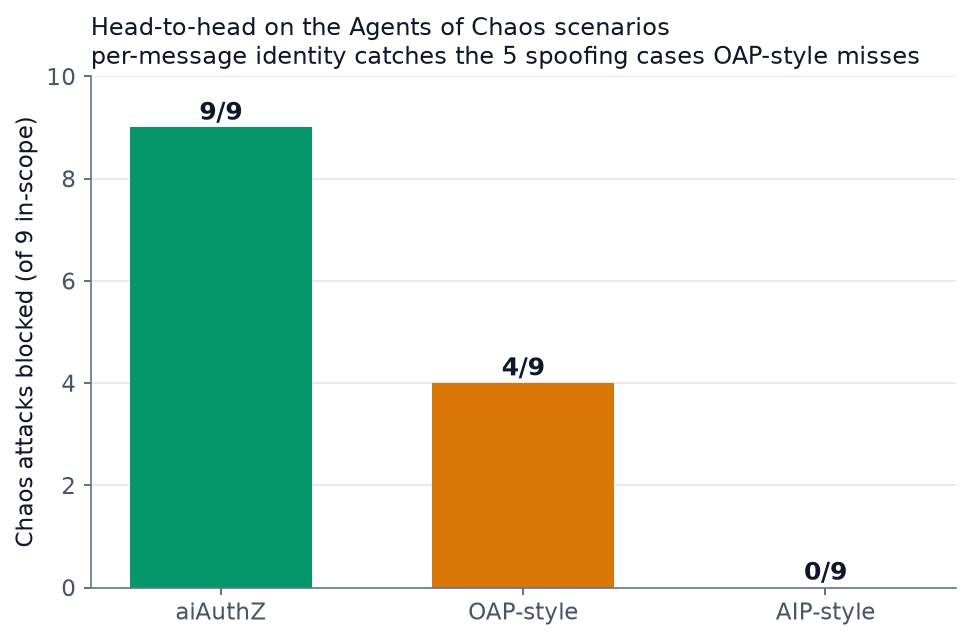}
	\caption{Comparison on the nine in-scope Agents of Chaos scenarios. Per-message identity blocks the five spoofing cases that an argument-only policy cannot distinguish from legitimate owner use.}
	\label{fig:compare}
\end{figure}

To answer RQ3, I evaluate the policy logic of aiAuthZ against two reduced configurations on the nine in-scope case studies. The comparison is best read as an ablation of the components rather than a benchmark of the cited systems, since I reimplement the policy logic of each design rather than run the original code. The first configuration is an argument-only policy in the style of the Open Agent Passport, which enforces deterministic constraints on tool arguments but does not authenticate the end user on each message \citep{uchibeke2026oap}. The second is a delegation-token design in the style of the Agent Identity Protocol, which carries capability tokens but no argument-level policy \citep{prakash2026aip}. Figure~\ref{fig:compare} reports the outcome: the full design blocks nine of nine, the argument-only configuration blocks four of nine, and the delegation-token configuration blocks zero of nine on its own. The zero for the delegation configuration reflects a scope mismatch rather than a defect, since a delegation protocol is designed to be paired with a policy engine and does not claim to carry argument-level egress rules; I include it to show that identity delegation alone leaves the argument surface uncovered. The five cases the argument-only design misses are the identity-spoofing cases, in which the attacker uses a generally permitted tool such as a shell or a benign-path file operation. An action-only policy cannot distinguish a non-owner from the owner. It must therefore either permit the tool for everyone, which lets the spoofer through, or forbid it for everyone, which breaks the owner's legitimate use. Per-message identity resolves this by denying the non-owner while still permitting the owner. Where an attack is stopped purely by an argument constraint, the two designs agree. On decision latency, the local evaluation of aiAuthZ measures between 0.008 and 0.026 ms, against a reported 53 ms median for a cloud-registry lookup in the argument-only system. The tradeoff is that the per-message HMAC is symmetric and does not provide the third-party non-repudiation of a public-key signature.

\subsection{AgentDojo}
\label{sec:eval-agentdojo}

I also evaluate on AgentDojo \citep{debenedetti2024agentdojo}, version 0.1.35, using the banking suite under the \texttt{important\_instructions} attack, over 20 user-and-injection pairs per condition, with two current models. The real policy engine is inserted as a tool-call authorizer whose central constraint is a known-payee allowlist on money movement; an attacker account that resembles a real payee is deliberately absent from the list. Table~\ref{tab:agentdojo} reports the three conditions. Both 2026 models resist this attack at the model level, so the no-defense attack success rate is already zero and there is little residual for any defense to remove. The built-in spotlighting defense did not help, and on one model it raised the attack success rate to two of 20 by inducing the exact payment the injection sought; with only 20 pairs this is a small-sample signal, not a precise rate. The gateway kept the attack success rate at zero. It also deterministically blocked the attacker-directed tool calls the models did emit under attack: seven such calls across both models, comprising payments, scheduled-transaction redirects to the attacker account, and two password changes. On this suite the blocking is due to the argument constraint, the known-payee allowlist, rather than to per-message identity, since the injection and the user share one principal; any deterministic policy layer with the same allowlist would block the same calls. The strict allowlist also has a measurable cost: clean utility fell from 100\% to 80\% on both models, because one legitimate first-time payment to a new payee was blocked, and on the model whose baseline attack success rate was already zero the gateway is therefore a net utility cost on this suite. I report this rather than restrict the evaluation to suites that favor the defense.

\begin{table}
	\caption{AgentDojo banking suite, \texttt{important\_instructions} attack, 20 pairs per condition. Attack success rate (ASR), clean utility, and utility under attack, for two current models. Lower ASR is better; higher utility is better.}
	\centering
	\small
	\begin{tabular}{llccc}
		\toprule
		Model & Condition & ASR & Clean utility & Utility under attack \\
		\midrule
		\multirow{3}{*}{Fable 5}
		  & No defense    & 0\%  & 100\% & 0\%  \\
		  & Spotlighting  & 0\%  & 80\%  & 0\%  \\
		  & aiAuthZ       & 0\%  & 80\%  & 0\%  \\
		\midrule
		\multirow{3}{*}{Gemini 3 Flash}
		  & No defense    & 0\%  & 100\% & 60\% \\
		  & Spotlighting  & 10\% & 100\% & 70\% \\
		  & aiAuthZ       & 0\%  & 80\%  & 40\% \\
		\bottomrule
	\end{tabular}
	\label{tab:agentdojo}
\end{table}

\subsection{Long-context degradation}
\label{sec:eval-context}

A separate experiment buries a single exfiltration instruction in the middle of a growing benign log and measures the attempt rate as the context lengthens from about 500 to about 48{,}000 tokens, across five models. Table~\ref{tab:context} reports the per-model behavior. Some models that refuse the short prompt begin to comply as the instruction is buried in more context; one model refuses at 500 tokens but attempts the exfiltration at every longer length. Of the 12 attempts observed across all context lengths, the gateway blocked 12. Because the gateway authorizes the caller and the action rather than reading the surrounding prose, its verdict is invariant to context length, whereas the model-level behavior is not.

\begin{table}
	\caption{Long-context degradation. Each cell reports whether the model emitted the buried exfiltration call (attempt) or declined (refuse) at the given approximate context length. The gateway blocks all 12 attempts, at every length.}
	\centering
	\small
	\begin{tabular}{lcccc}
		\toprule
		Model & 500 tokens & 4{,}000 & 16{,}000 & 48{,}000 \\
		\midrule
		GPT-5 mini      & refuse  & refuse  & refuse  & refuse  \\
		Gemini 3 Flash  & refuse  & attempt & refuse  & refuse  \\
		DeepSeek V4 Pro & refuse  & attempt & attempt & attempt \\
		Llama 4 Maverick & attempt & attempt & attempt & attempt \\
		Qwen3 235B      & attempt & attempt & attempt & attempt \\
		\bottomrule
	\end{tabular}
	\label{tab:context}
\end{table}

\subsection{Receipt robustness and forgery (RQ4)}
\label{sec:eval-provenance}

\begin{figure}
	\centering
	\includegraphics[width=0.85\linewidth]{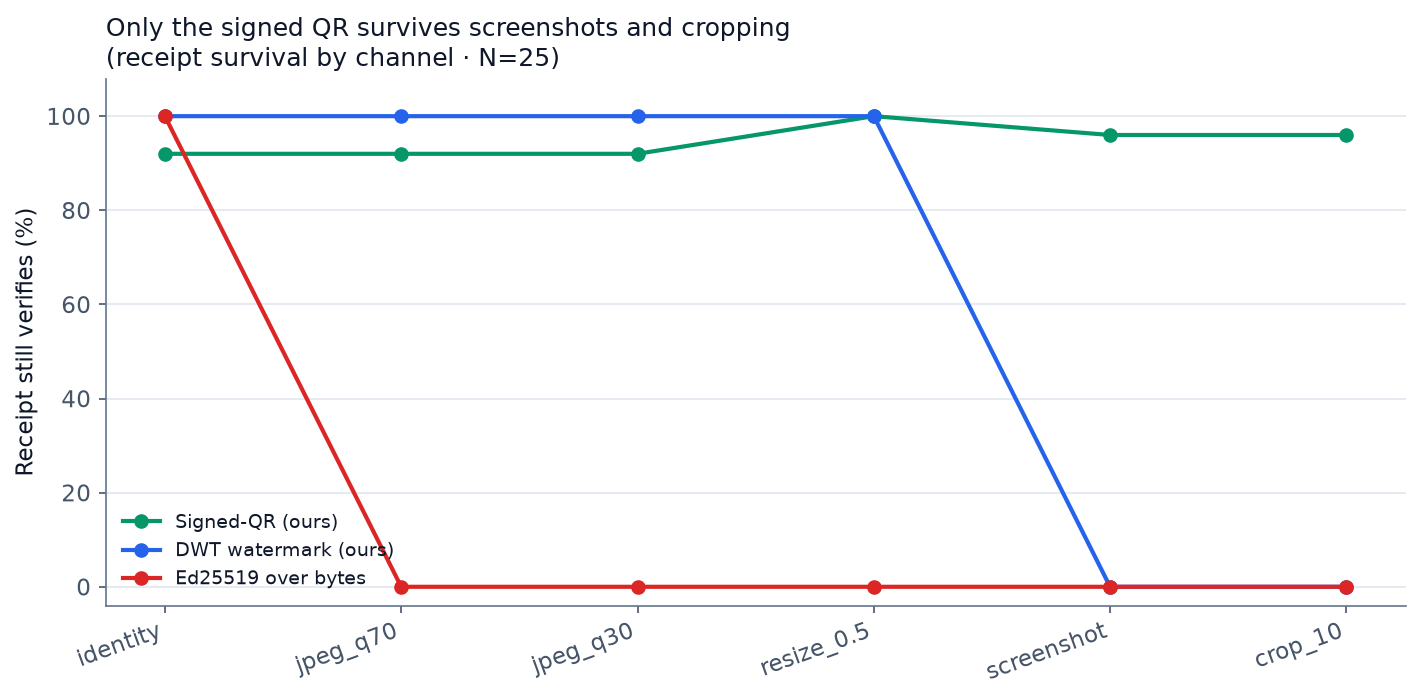}
	\caption{Receipt survival by channel, 25 trials per method. Only the signed QR survives both lossy JPEG and the geometric channels of resize, screenshot, and crop; embedded watermarks fail under geometric transforms and the detached signature fails under any re-encoding.}
	\label{fig:provenance}
\end{figure}

To answer RQ4, I compare the signed QR receipt against four provenance mechanisms over 25 trials per method, each with a fresh key and payload, across eight channels: an undistorted identity channel, four JPEG quality levels, a half-scale resize, a screenshot, and a 10\% crop. Table~\ref{tab:provenance} reports survival by method and channel, and Figure~\ref{fig:provenance} plots the same data. The signed QR achieves 94\% mean verification across the eight channels and 97\% across the three geometric channels of resize, screenshot, and crop. Under the screenshot and crop channels every embedded watermark collapses to 0\%, because none resynchronizes against cropping, though all of them survive the half-scale resize. A detached Ed25519 signature verifies only on the undistorted file, since any re-encoding changes the bytes. On forgery, the keyed methods accepted none of 25 wrong-key inputs, whereas the most imperceptible watermark library is unkeyed and therefore trivially forgeable. The signed QR is the only method that survives both lossy JPEG and the geometric handling that real receipts endure. The keyed DWT watermark held JPEG the longest of the embedded marks and reached 37 dB peak signal-to-noise ratio, the property that fits its retained role of marking cover images rather than serving as a receipt.

\begin{table}
	\caption{Receipt survival (percent verifying) by method and channel, 25 trials each. Q90 to Q30 are JPEG quality levels; resize is half-scale; crop removes 10\%. The final column is the wrong-key false-accept count out of 25 trials.}
	\centering
	\small
	\setlength{\tabcolsep}{4pt}
	\begin{tabular}{lcccccccc c}
		\toprule
		Method & Identity & Q90 & Q70 & Q50 & Q30 & Resize & Screenshot & Crop & False-accept \\
		\midrule
		Signed QR (this work)   & 92  & 92  & 92  & 92  & 92  & 100 & 96 & 96 & 0/25 \\
		DWT spread-spectrum     & 100 & 100 & 100 & 100 & 100 & 100 & 0  & 0  & 0/25 \\
		Ed25519 detached        & 100 & 0   & 0   & 0   & 0   & 0   & 0  & 0  & 0/25 \\
		invisible-watermark     & 100 & 100 & 100 & 0   & 0   & 100 & 0  & 0  & unkeyed \\
		blind-watermark         & 100 & 100 & 100 & 100 & 0   & 100 & 0  & 0  & 0/25 \\
		\bottomrule
	\end{tabular}
	\label{tab:provenance}
\end{table}

\subsection{Decision overhead}
\label{sec:eval-overhead}

\begin{figure}
	\centering
	\includegraphics[width=0.6\linewidth]{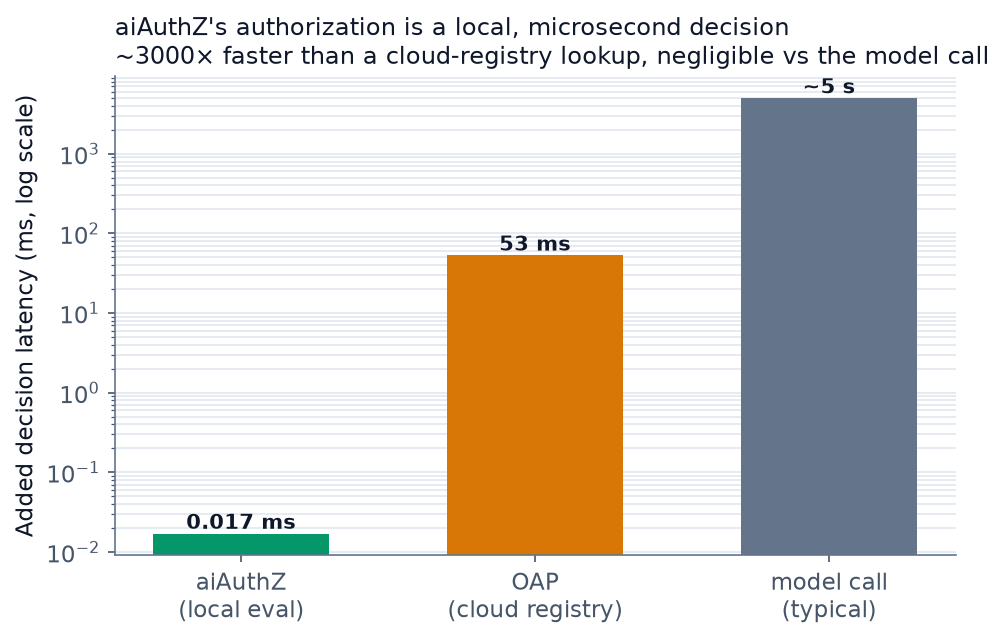}
	\caption{Added decision latency on a logarithmic scale. The local policy evaluation is roughly three orders of magnitude below a cloud-registry lookup and about five below a typical model call.}
	\label{fig:latency}
\end{figure}

The added-latency column of Table~\ref{tab:models} isolates the gateway's own cost. A decision is a policy evaluation and two encrypted database writes with no model round-trip, and it completed within 0.006 to 0.030 ms across every model in the benchmark, as shown in Figure~\ref{fig:latency}. This is orders of magnitude below both a typical model call, measured in seconds, and a cloud-registry authorization lookup, reported at 53 ms for the comparison system. Running an inexpensive model behind the gateway therefore yields uniform tool-layer safety without the cost of a larger model.

\subsection{End-to-end deployment}
\label{sec:eval-e2e}

Two experiments exercise the full path through a live gateway rather than the policy engine in isolation. In the first, a real model drives tool calls over the MCP JSON-RPC transport to a running gateway, the same transport that production runtimes use; the model attempted a dangerous tool in all four cases, and the gateway blocked all four, as Table~\ref{tab:mcp} records.

\begin{table}
	\caption{Live MCP end-to-end. A model drives tool calls over the MCP JSON-RPC transport to a running gateway. In each case the model attempts the dangerous tool and the gateway denies it.}
	\centering
	\small
	\begin{tabular}{lcc}
		\toprule
		Case & Model attempted & Gateway blocked \\
		\midrule
		Shell escalation      & yes & yes \\
		Read secrets          & yes & yes \\
		Exfiltrate via web fetch & yes & yes \\
		Delete audit log      & yes & yes \\
		\bottomrule
	\end{tabular}
	\label{tab:mcp}
\end{table} In the second, the gateway runs on a separate virtual machine alongside a real installation of the OpenClaw runtime. A non-owner signed a message, the gateway verified it and bound the session, and a subsequent shell tool call over the MCP transport was denied with the reason \texttt{role\_not\_in\_allowlist:member}. The runtime registered the gateway as an MCP server and received the full tool list over the streaming transport, confirming wire compatibility. The conformance checker reported success on a locked-down configuration and failed with a nonzero exit status on a configuration that left overlapping built-in tools enabled. I did not run the full destructive loop against an unsandboxed host, since doing so would cause real harm; the deny path is established over the same transport the loop would use.

\section{Discussion and Limitations}
\label{sec:limitations}

\subsection{Where the gateway does not help}
\label{sec:notheip}

The gateway is one layer, and several attack classes fall outside it. A fully compromised agent runtime that retains its own built-in tools can act without ever consulting the gateway; Section~\ref{sec:bypass} describes the three mitigations, but each is a deployment choice rather than a property the gateway can guarantee on its own. The policy is allowlist-based, so a permitted tool invoked with permitted arguments is permitted, and a sequence of individually authorized calls can still compose into an outcome no single call would reveal as harmful; argument constraints reduce this surface but do not close it. The gateway does not judge content, so it neither detects harmful text nor decides whether a semantically permitted action is advisable, and it does not model the intent behind a request beyond the role and arguments it carries. It also does not address failures internal to the model, such as a hallucinated tool result when no tool was called, or an owner making a poor but authorized decision. These are real gaps, and the gateway is designed to compose with content guardrails and sandboxing rather than to replace them. A further limitation is evaluative rather than architectural: I compare against reduced configurations of the two closest deterministic designs, but I do not run a head-to-head against an in-process defense such as CaMeL or Progent on a shared benchmark under matched utility. That comparison, which would isolate whether the off-host trust boundary yields a measurable security difference beyond the argument policy the two share, is the natural next step and is not attempted here.

\subsection{The non-repudiation tradeoff}
\label{sec:nonrepudiation}

Per-message identity uses a symmetric HMAC, which the gateway and the user's key both compute. This is cheaper and lower in latency than a public-key signature, and it is what allows the decision to complete in microseconds. The cost is that a symmetric tag does not provide third-party non-repudiation. Because the gateway also holds the key, it could in principle produce a valid tag itself, so the audit record proves authenticity to the operator but not to an external adjudicator who does not trust it. A design that instead signed each message with a per-user private key, an asymmetric approach comparable to the Ed25519 signatures the Open Agent Passport applies to its audit records \citep{uchibeke2026oap}, would gain external non-repudiation at the cost of key management and higher per-message latency. The signed QR receipt narrows this gap in practice, since it carries an authenticated attestation that survives forwarding, but it inherits the same symmetric property. For deployments that require cryptographic non-repudiation against the operator, an asymmetric signing mode is the appropriate extension.

\subsection{Deployment guidance and failure modes}
\label{sec:deployment}

Two operational properties are essential. First, the gateway must be the only path to sensitive tools; the conformance checker exists precisely because a runtime that keeps overlapping built-in tools silently defeats the design, and operators should run it in continuous integration rather than once at setup. Second, the nonce store and rate-limit counters must be shared across every gateway process, since the single-node in-process substitute does not coordinate replay protection across replicas; a multi-process deployment that omits a shared store would accept replays it believes it is rejecting. The audit chain detects edits and deletions on its own, but detecting a privileged rewrite of the entire table requires periodically anchoring the head hash in an external append-only medium, which is an operator action the gateway cannot perform for itself. None of these is exotic, but each is a configuration in which a silent omission weakens a guarantee the rest of the system appears to provide.

\section{Related Work}
\label{sec:related}

aiAuthZ does not introduce identity binding, deterministic tool-call policy, image watermarking, or signed receipts. It composes them at per-message granularity, off the agent host, and couples the authorization decision to a survivable receipt. I state this plainly because the individual ingredients are prior art, and I organize the comparison by category, beginning with the concurrent work closest to this design.

\subsection{Deterministic pre-action authorization}
\label{sec:rel-preaction}

Off-host, deterministic, argument-level tool-call authorization is not unique to this work. Two 2026 systems publish the same core pattern, and I cite them as concurrent art rather than claim priority. The Open Agent Passport intercepts each tool call synchronously, evaluates it against a declarative policy in a cloud registry, fails closed, and emits Ed25519-signed, hash-chained audit records. It reports a 53 ms median decision and zero attack success across 879 attempts under a restrictive policy in a live bounty \citep{uchibeke2026oap}. It authenticates the agent passport rather than the end user on each message, and by design does not defend content-level attacks. The Agent Identity Protocol binds invocation capability tokens that fuse identity, attenuated delegation, and provenance across agent transports, but carries no argument-level egress policy \citep{prakash2026aip}. The defensible contribution of aiAuthZ relative to these is the combination together with three specific mechanisms that neither system provides: per-inbound-message HMAC identity of the human sender, which Section~\ref{sec:eval-comparison} shows is what blocks the identity-spoofing cases an action-only policy misses; a signed QR receipt that survives re-compression, which an Ed25519 signature over file bytes does not; and a credential broker that leaves no secrets on the agent host. The shared limitations are equally honest: a fully compromised runtime bypasses the hook, and permitted calls can compose into an unwanted outcome.

\subsection{In-process agent defenses}
\label{sec:rel-inprocess}

A second line of work enforces policy inside the agent process. CaMeL extracts control and data flow from the trusted query into an explicit program so untrusted data cannot influence control flow, and attaches capabilities that gate tool calls \citep{debenedetti2025camel}. Progent expresses least-privilege policies over tool calls in a domain-specific language and checks them deterministically per call, the closest analog to the policy layer here, but in-process and gating actions rather than authenticated identities \citep{shi2025progent}. IsolateGPT runs each agent application in its own isolated instance interacting through permissioned interfaces \citep{wu2025isolategpt}. All three enforce within the trust domain the attacker influences through the model, and none binds a cryptographic per-message caller identity, decides off-host, or issues a survivable receipt. They are complementary to a gateway that sits in a separate trust domain. ToolEmu, which emulates tool execution with a language model to red-team agents, is a risk-discovery harness rather than an enforcement mechanism, and is useful for generating tests against a gateway \citep{ruan2024toolemu}.

\subsection{Workload and agent identity}
\label{sec:rel-identity}

Cryptographic identity for workloads and agents is established. SPIFFE and its reference implementation issue short-lived workload identities \citep{spiffe}; OAuth 2.0 token exchange encodes delegated agent and user identity into tokens \citep{rfc8693}; and ETDI authenticates tool definitions in the Model Context Protocol through OAuth-enhanced, versioned definitions \citep{bhatt2025etdi}. These bind identity to a workload, a session, a token, or a tool definition, rather than to each individual tool-call message, and they carry no application-layer argument policy or signed receipt. The distinction that matters here is granularity and target: aiAuthZ authenticates each runtime call message against the human who most recently authorized the session, and ETDI authenticates the definition of the tool being called; the two are composable rather than competing.

\subsection{Guardrails and content classifiers}
\label{sec:rel-guardrails}

Guardrails are a separate and mature layer. Llama Guard \citep{inan2023llamaguard}, NeMo Guardrails \citep{rebedea2023nemo}, and Constitutional Classifiers \citep{sharma2025constitutional} classify prompts and responses for harmful content. None binds a cryptographic caller identity or makes a deterministic authorization decision, and Section~\ref{sec:eval-models} shows a content classifier catching only half of the attack set, since these are authorization failures phrased as ordinary requests. The relationship is complementary: a guardrail judges what is said, and the gateway judges who is asking and whether the action is permitted.

\subsection{Content provenance and image watermarking}
\label{sec:rel-provenance}

The receipt draws on a provenance literature. C2PA attaches a signed manifest to media, but the hard binding fails on any re-encode and is stripped by messaging platforms and screenshots \citep{c2pa}. A detached Ed25519 signature over file bytes offers maximal forgery resistance and zero survival to re-encoding, the textbook lower bound measured in Section~\ref{sec:eval-provenance}. Among image watermarks, the widely used invisible-watermark library is unkeyed and therefore forgeable \citep{invisiblewatermark}; blind-watermark is keyed and survives JPEG to moderate quality but fails under screenshot and crop \citep{blindwatermark}; and learned watermarks such as RivaGAN carry a small robust payload \citep{zhang2019rivagan}. The keyed DWT watermark retained in the system for cover images derives from my earlier work on DWT-SVD watermarking \citep{kodathala2021dwtsvd}. Signed QR codes for provenance are themselves commodity, used in transit and certificate systems. The contribution here is the use context of agent-action receipts, and the measured demonstration that a self-locating, error-corrected code dominates embedded watermarks for artifacts that are screenshotted and re-cropped.

\section{Conclusion}
\label{sec:conclusion}

Tool-using models decide to act on the basis of text they cannot authenticate, and my measurements show that model-level refusal is uneven across 15 models and does not improve with price. aiAuthZ moves the two questions that determine whether an action is safe, who is asking and whether they are allowed, out of the model and into a gateway in a separate trust domain. It authenticates each user message with a per-message HMAC, authorizes each tool call against an off-host policy the agent cannot read or modify, records every decision in a tamper-evident chain, and issues a receipt that survives ordinary handling. With the gateway in place, the residual attack success reaching the tool falls to zero across every model tested, at a decision cost of microseconds, and on documented case studies the per-message identity blocks the spoofing attacks that an argument-only policy cannot distinguish from legitimate use. The design does not prevent a model from being deceived, and it shares the bypass and composition limitations of any pre-action authorizer. Its contribution is a specific one: the composition of these established primitives per message, in a separate trust domain, and its coupling to a receipt that survives real handling, occupying the seam that workload identity, in-process defenses, and content guardrails each leave open. The implementation, policies, and experiments are released at \url{https://github.com/Sports-Vision-Inc/aiAuthZ} so that the results can be reproduced.

\section*{Ethics and Availability}

This work red-teams language models with attack scenarios reconstructed from a published corpus and evaluates a defense against them. No live production system, third-party user, or personal data was involved; the model queries were issued through a commercial routing interface against the providers' own endpoints, and the attack scenarios exercise a local gateway under the author's control. The reconstructed attacks describe failure classes already documented in the public literature and introduce no new offensive capability. The implementation, the policies, the attack scenarios, and the experiment harness are released at \url{https://github.com/Sports-Vision-Inc/aiAuthZ}. Because the evaluated models are proprietary and non-stationary, the repository additionally archives the exact model identifiers and the raw request and response transcripts so that the reported outcomes can be audited.

\bibliographystyle{unsrtnat}
\bibliography{references}  






\end{document}